\documentclass[prl,aps,nofootinbib,twocolumn]{revtex4}%,%onecolumn,%groupedaddress]{revtex4}
\usepackage{epsfig}
\usepackage{amsfonts}
\usepackage{rotating}
\usepackage{sparticles}
\usepackage{dcolumn}
\usepackage{bm}
\usepackage{hyperref}
\hypersetup{
    bookmarks=true,         % show bookmarks bar?
    unicode=false,          % non-Latin characters in Acrobat’s bookmarks
    pdftoolbar=true,        % show Acrobat’s toolbar?
    pdfmenubar=true,        % show Acrobat’s menu?
    pdffitwindow=false,     % window fit to page when opened
    pdfstartview={FitH},    % fits the width of the page to the window
    pdftitle={My title},    % title
    pdfauthor={Author},     % author
    pdfsubject={Subject},   % subject of the document
    pdfcreator={Creator},   % creator of the document
    pdfproducer={Producer}, % producer of the document
    pdfkeywords={keyword1} {key2} {key3}, % list of keywords
    pdfnewwindow=true,      % links in new window
    colorlinks=true,       % false: boxed links; true: colored links
    linkcolor=red,          % color of internal links
    citecolor=cyan,        % color of links to bibliography
    filecolor=magenta,      % color of file links
    urlcolor=green,           % color of external links
    linktocpage=true
}
\usepackage{amsmath,amssymb,amsthm,graphicx,latexsym}
\usepackage{subfigure}
\usepackage{pstricks}
\usepackage{color}

\usepackage{mathrsfs}

\newcommand{\be}{\begin{equation}}
\newcommand{\ee}{\end{equation}}
\newcommand{\bea}{\begin{eqnarray}}
\newcommand{\eea}{\end{eqnarray}}
\newcommand{\bml}{\begin{subequations}}
\newcommand{\eml}{\end{subequations}}
\newcommand{\bfig}{\begin{figure}}
\newcommand{\efig}{\end{figure}}

\newcommand{\slashed}{\hspace{-1.1ex}/}

\begin{document}

\title{Sub-Planckian inflation \& large tensor to scalar ratio with $r\geq 0.1$}
\author{Sayantan Choudhury $^{1}$ and Anupam Mazumdar $^{2}$}
\affiliation{$^1$Physics and Applied Mathematics Unit, Indian Statistical Institute, 203 B.T. Road, Kolkata 700 108, INDIA}
\affiliation{$^2$Consortium for Fundamental Physics, Physics Department, Lancaster University, LA1 4YB, UK
}

%\vspace{5ex}
%\date{\today}
\begin{abstract}
We categorically point out why the analysis of Ref.~\cite{Antusch:2014cpa} is incorrect. Here we 
explicitly show why the sub-Planckian field excursion of the inflaton field can yield large observable tensor-to-scalar ratio, which satisfies 
both Planck and BICEP constraints.
\end{abstract}

%\pacs{98.80.-k ; 98.80.Cq ; 04.50.-h}

\maketitle

We have shown in Refs.~\cite{Choudhury:2013iaa} and \cite{Choudhury:2014kma} that the sub-Planckian excursion of the inflaton field can generate large value of tensor-to-scalar
ratio as observed by BICEP2 and also satisfies the constraints obtained from the Planck after foreground subtractions~\footnote{An explicit example was provided earlier in the 
context of high scale MSSM inflation in presence of Hubble induced K\"{a}hler corrections in Ref.~\cite{Choudhury:2013jya}.}. 
However, recently it was claimed in Ref.~\cite{Antusch:2014cpa} that for a single field inflationary model with sub-Planckian field excursion it is not possible to
generate the observed large tensor-to-scalar ratio. Unfortunately, the validity of this claim is completely wrong. In this short report our prime objective is to 
explicitly show why the claim in  Ref.~\cite{Antusch:2014cpa} is wrong while providing explicitly the steps which the authors completely ignored.

Here we will refute the points raised in Ref.~\cite{Antusch:2014cpa}, while clarifying the analytics explicitly:
\begin{itemize}
 \item \underline{\bf Step 1}: In Refs.~\cite{Choudhury:2013iaa,Choudhury:2014kma}, we considered a generic potential, which is expanded in a Taylor series around the sub-Planckian 
VEV, $\phi_{0}<M_{P}$ as:
  \be\begin{array}{llll}\label{eq1}
  \displaystyle    V(\phi)=V(\phi_0)+V^{'}(\phi_0)(\phi-\phi_0)+\frac{V^{''}(\phi_0)}{2}(\phi-\phi_0)^{2}\\
~~~~~~~~~~~~~~~~~~+\displaystyle \frac{V^{'''}(\phi_0)}{6}(\phi-\phi_0)^{3}+\frac{V^{''''}(\phi_0)}{24}(\phi-\phi_0)^{4}
     \end{array}\ee
where we have truncated the Taylor expansion as: $V(\phi_0)>V^{'}(\phi_0)>V^{''}(\phi_0)>V^{''''}(\phi_0)$ (in the Planckian unit)
, which is also the necessary condition for the convergence of the Taylor series. Note that $\phi_0$ denotes the VEV where inflation occurs in its vicinity.

\item \underline{\bf Step 2}:
We can derive a simple expression for the tensor-to-scalar ratio, $r$, as, see~\cite{Choudhury:2013jya,Choudhury:2013iaa,Choudhury:2014kma}:
\be\label{eq2}
 r=\frac{8}{M^{2}_{p}}\frac{(1-\epsilon_{V})^{2}\left[1-({\cal C}_{E}+1)\epsilon_{V}\right]^{2}}{\left[1-(3{\cal C}_{E}+1)\epsilon_{V}
+{\cal C}_{E}\eta_{V}\right]^{2}}\left(\frac{d\phi}{d{\ln k}}\right)^{2}+\cdots\,,
 \ee
where ${\cal C}_{E}=4(\ln 2+\gamma_{E})-5$ with $\gamma_{E}=0.5772$ is the {\it Euler-Mascheroni constant}, $\epsilon_V,~\eta_V$
are slow roll parameters~\footnote{We use the standard notations and for details readers can see Refs.~\cite{Choudhury:2013iaa,Choudhury:2014kma}.},
there are 
higher order terms in slow roll parameters, of order ${\cal O}(\epsilon ^2_{V}),~{\cal O}(\eta^2_{V})\cdots$, which will give negligible contributions and would not alter 
the results of our discussion. We can now derive a bound on $r(k)$  in terms of the momentum scale:
\be\begin{array}{llll}\label{eq3}
    \displaystyle \left|\int^{{ k}_{\star}}_{{k}_{e}}\frac{dk}{k}\sqrt{\frac{r({k})}{8}} \right|
\displaystyle \approx \frac{|\Delta\phi|}{M_p} \left\{ 1+\underbrace{....}_{<<1}\right\}\approx \frac{|\Delta\phi|}{M_p}\,,
   \end{array}\ee
where $\Delta\phi=\phi_{\star}-\phi_{e}$ and we have neglected the contributions from the $\underbrace{....}$ terms as they are small compared to the leading order term
due to the convergence of the series mentioned in Eq~(\ref{eq1}). Here $\phi_e$ denotes the inflaton VEV at the end of inflation,
and $\phi_{\star}$ denote the field VEV when the corresponding mode $k_\star$ is leaving the Hubble patch during inflation.

\item \underline{\bf Step 3}: 
In order to perform the momentum integration in the left hand side of Eq~(\ref{eq3}) analytically, we have used the following parameterization of
 $r(k)$, which can be expressed as~\footnote{Note that in the following expression, Eq.~(\ref{eq4}), we have taken running and running of the spectrum, while in
 Eq.~(\ref{eq2}) we have only taken the leading order contribution which mainly involves $\epsilon_V,~\eta_V$. The procedure is perfectly correct, since the higher order 
 corrections are sub-leading. This is precisely by virtue of the Taylor expansion of the potential in the vicinity of $\phi_0$ where inflation occurs.}:
\be\label{eq4}
 r(k)=r(k_{\star})\left(\frac{k}{k_{\star}}\right)^{a+\frac{b}{2}\ln\left(\frac{k}{k_{\star}}\right)
+\frac{c}{6}\ln^{2}\left(\frac{k}{k_{\star}}\right)}\,,
\ee
where 
\be\label{eq5}
a=n_{T}-n_{S}+1,~~~b=\left(\alpha_{T}-\alpha_{S}\right),~~~c=\left(\kappa_{T}-\kappa_{S}\right)\,.
\ee
which are defined at the scale $k_{\star}$.
These parameterization characterizes the spectral indices, $n_S,~n_T$, running 
of the spectral indices, $\alpha_S,~\alpha_T$, and running of the running of the spectral indices, $\kappa_S,~\kappa_T$. 
Here the subscripts, $(S,~T)$, represent the scalar and tensor modes.

After substituting Eq~(\ref{eq4}) in the left hand side of we Eq~(\ref{eq3}), we obtain:
\be\begin{array}{llll}\label{eq6}
    \displaystyle \int^{{ k}_{\star}}_{{k}_{e}}\frac{dk}{k}\sqrt{\frac{r({k})}{8}}\\
\displaystyle =\sqrt{\frac{r(k_{\star})}{8}}\int^{{ k}_{\star}}_{{k}_{e}}\frac{dk}{k}\sqrt{\left(\frac{k}{k_{\star}}\right)^{a+\frac{b}{2}\ln\left(\frac{k}{k_{\star}}\right)
+\frac{c}{6}\ln^{2}\left(\frac{k}{k_{\star}}\right)}},\\
\displaystyle =\sqrt{\frac{r(k_{\star})}{8}}\int^{{ k}_{\star}}_{{k}_{e}}\frac{dk}{k_{\star}}\left(\frac{k}{k_{\star}}\right)^{A+B\ln\left(\frac{k}{k_{\star}}\right)
+C\ln^{2}\left(\frac{k}{k_{\star}}\right)},
   \end{array}\ee
where $$A=\left(\frac{a}{2}-1\right),~B=\frac{b}{4},~C=\frac{c}{12}.$$

Let us substitute, $k/k_{\star}=\ln y$, to simplify the mathematical form of the above Eq~(\ref{eq6}). Consequently, we get:
\be\begin{array}{llll}\label{eq7}
    \displaystyle \int^{{ k}_{\star}}_{{k}_{e}}\frac{dk}{k}\sqrt{\frac{r({k})}{8}}\\
\displaystyle =\sqrt{\frac{r(k_{\star})}{8}}\int^{e^{1}}_{e^{{k}_{e}/k_{\star}}}\frac{dy}{y}\left(\ln y\right)^{A+B\ln\left(\ln y\right)
+C\ln^{2}\left(\ln y\right)},\end{array}\ee
To evaluate the integral analytically, we apply the following technique.
Let us consider:
\be\begin{array}{lll}\label{f1}
(\ln y)^{\alpha},~~~ {\rm where} ~~~\alpha<<1
\end{array}\ee
where the exponent $\alpha$ is defined as:
\be\label{f2}
\alpha=A+B\ln\left(\ln y\right)
+C\ln^{2}\left(\ln y\right)
\ee
where $|A|,|B|,|C| \ll 1$ with $|A|>|B|>|C|$. Now, for $\alpha<<1$, which is  typically the case, one can expand the 
function mentioned in Eq~(\ref{f1}) as~\footnote{One can verify that $\alpha << 1$ for a slow roll inflation, within the interval $8.2\times 10^{-11}~{\rm Mpc}^{-1}
\leq k\leq 0.056~{\rm Mpc}^{-1}$.}: 
\be\begin{array}{lll}\label{f3}
(\ln y)^{\alpha}=1+\alpha\ln(\ln y)+\cdots
\end{array}\ee

 Let us take first two terms in the right hand side of the series expansion.
This finally results in:
\be\begin{array}{llll}\label{eq7a}
\displaystyle \int^{{ k}_{\star}}_{{k}_{e}}\frac{dk}{k}\sqrt{\frac{r({k})}{8}}\\
\displaystyle \approx\sqrt{\frac{r(k_{\star})}{8}}\int^{e^{1}}_{e^{{k}_{e}/k_{\star}}}\frac{dy}{y}\left\{1+\left[A+B\ln\left(\ln y\right)
\right.\right.\\ \left.\left.~~~~~~~~~~~~~~~~~~~~~~~~~~~~~~~~\displaystyle +C\ln^{2}\left(\ln y\right)\right]\ln (\ln y)\right\},\\
\displaystyle =\sqrt{\frac{r(k_{\star})}{8}}\left[\left(1-A+2B-6C\right)\ln y
\right.\\ \left.\displaystyle ~~~~~~~~~~~~~~~~~ +\left(A-2B+6C\right)(\ln y)\ln(\ln y) 
\right.\\ \left.\displaystyle ~~~~~~~~~~~~~~~~~ +\left(B-3C\right)(\ln y)\ln^{2}(\ln y) 
\right.\\ \left.\displaystyle ~~~~~~~~~~~~~~~~~~~~~~~~~~~~~~~~~~~~+C(\ln y)(\ln(\ln y))^{3}\right]^{e^{1}}_{e^{{k}_{e}/k_{\star}}},\\
\displaystyle =\sqrt{\frac{r(k_{\star})}{8}}\left[\left(1-A+2B-6C\right)\left[1-\frac{k_{e}}{k_{\star}}\right]
\right.\\ \left.\displaystyle ~~~~~~~~~~~~~~~~~ -\left(A-2B+6C\right)\frac{k_{e}}{k_{\star}}\ln\left(\frac{k_{e}}{k_{\star}}\right) 
\right.\\ \left.\displaystyle ~~~~~~~~~~~~~~~~~ -\left(B-3C\right)\frac{k_{e}}{k_{\star}}\ln^{2}\left(\frac{k_{e}}{k_{\star}}\right) 
\right.\\ \left.\displaystyle ~~~~~~~~~~~~~~~~~~~~~~~~~~~~~~~~~~~~-C\frac{k_{e}}{k_{\star}}\ln^{3}\left(\frac{k_{e}}{k_{\star}}\right)\right],\\
\\
\displaystyle =\sqrt{\frac{r(k_{\star})}{8}}\left[\left(2-\frac{a}{2}+\frac{b}{2}-\frac{c}{2}\right)\left[1-\frac{k_{e}}{k_{\star}}\right]
\right.\\ \left.\displaystyle ~~~~~~~~~~~~~~~~~ -\left(\frac{a}{2}-\frac{b}{2}+\frac{c}{2}-1\right)\frac{k_{e}}{k_{\star}}\ln\left(\frac{k_{e}}{k_{\star}}\right) 
\right.\\ \left.\displaystyle ~~~~~~~~~~~~~~~~~ -\left(\frac{b}{4}-\frac{c}{4}\right)\frac{k_{e}}{k_{\star}}\ln^{2}\left(\frac{k_{e}}{k_{\star}}\right) 
\right.\\ \left.\displaystyle ~~~~~~~~~~~~~~~~~~~~~~~~~~~~~~~~~~~~-\frac{c}{12}\frac{k_{e}}{k_{\star}}\ln^{3}\left(\frac{k_{e}}{k_{\star}}\right)\right].
   \end{array}\ee

{\bf \red This is the first step where the  analysis done in  Ref.~\cite{Antusch:2014cpa}  is wrong and also misleading - the authors numerically approximate the 
integrals by substituting the values of $a,~b$ and $c$, which strictly speaking one should not do. 
Also in their criticism - they are talking the limits $a,~b,~c\rightarrow 0$, which is completely incorrect, this would mean, 
$$n_{T}\rightarrow n_{S}-1\,,~\alpha_{T}\rightarrow\alpha_{S}\,,~\kappa_{T}\rightarrow \kappa_{S},$$ and such a hypothetical  situation is not supported by 
any inflationary models {\it at least known to us}, as far as the recent observational evidences from BICEP2 and Planck are concerned. Certainly not within {\it inflection} point 
models of inflation with a sub-Planckian VEV of inflaton, as described in Refs.~\cite{Choudhury:2013jya,Choudhury:2013iaa}. }

In our current analysis and before, see Refs.~\cite{Choudhury:2013iaa,Choudhury:2014kma}, we have explored the possibility of $a,~b,~c\neq 0$, and furthermore contributions
from $b$ and $c$ are not negligible due to the presence of running of scalar spectral tilt $\alpha_{S}$, and running of the running of the tilt $\kappa_{S}$
when taken both BICEP2 and Planck data within $1.5\sigma$. In support of this statement we have plotted the behaviour of the scalar power spectrum $P_{S(k) }$, and the
number of e-foldings of inflation, $N(k)$ in fig~(\ref{fig:subfig1},~\ref{fig:subfig4}) within the observed muiltpole of Planck, i.e. $2<l<2500$. 
%%%%%%%%%%%%%%%%%%%%%%%%%%%%%%%FIGURE%%%%%%%%%%%%%%%%%%%%%%%%%%%%%%%%%%%%%%%%%%%%%%

\begin{figure*}[t]
\centering
\subfigure[]{
    \includegraphics[width=8.1cm, height=6.5cm] {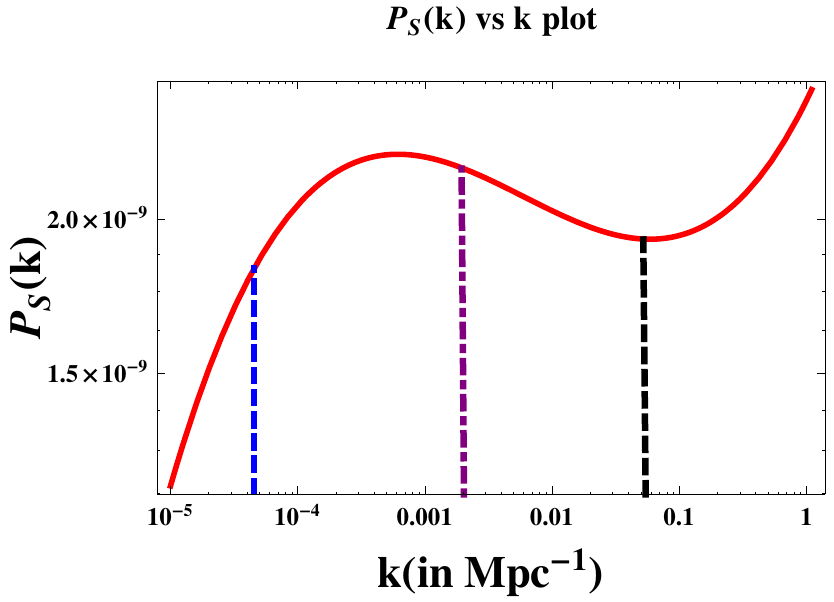}
    \label{fig:subfig1}
}
\subfigure[]{
    \includegraphics[width=8.1cm, height=6.5cm] {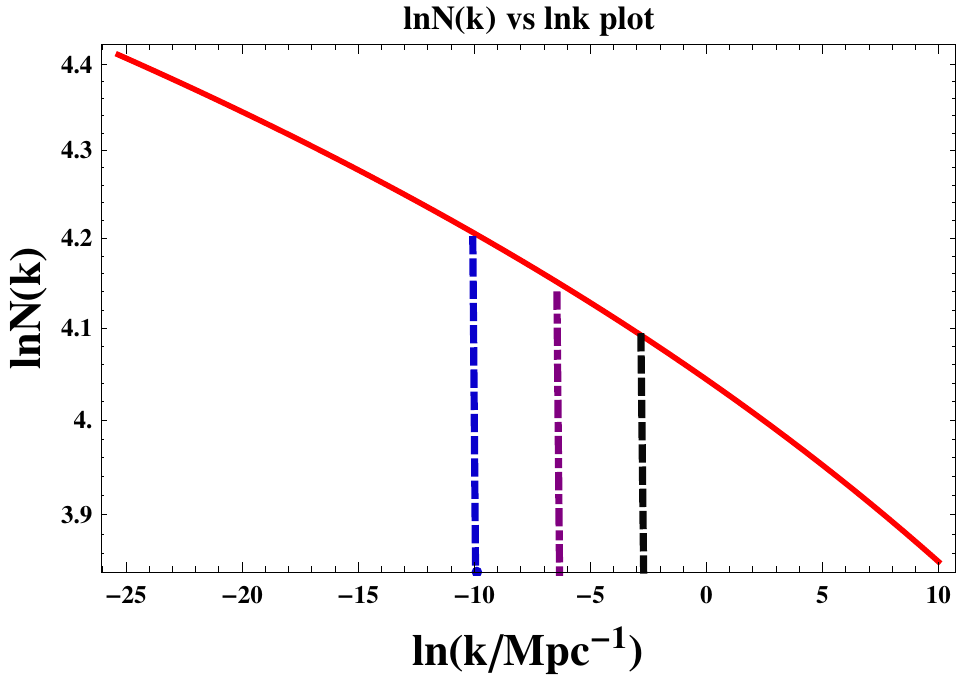}
    \label{fig:subfig4}
}
\caption[Optional caption for list of figures]{ In \ref{fig:subfig1}, we show the amplitude of the scalar power spectrum $P_{S}(k)$,
 and in \ref{fig:subfig4}, we show the total number of e-foldings $N(k)$, with respect to the momentum scale $k$.
The {\bf black} dotted line corresponds to $k_{max}=0.056~{\rm Mpc}^{-1}$ for $l_{max}=2500$, the \textcolor{blue}{\bf blue} dotted line corresponds to
$k_{min}=4.488\times 10^{-5}~{\rm Mpc}^{-1}$ for $l_{min}=2$,  and in all the plots \textcolor{violet}{\bf violet} dashed dotted line
 represents the pivot scale of momentum at $k_{\star}=0.002~{\rm Mpc}^{-1}$ for $l_{\star}\sim 80$ at which
 $P_{S}(k_{\star})=2.2\times 10^{-9}$, $n_{S}=0.96$, $\alpha_{S}=-0.02$ and $N(k_{\star})=63.26$. Within $2<l<2500$ the
 value of the required momentum scale is determined by the relation,
$k_{reqd}\sim \frac{l_{reqd}}{\eta_{0}\pi}$ \cite{Choudhury:2013jya}, where the conformal time at the present epoch is $\eta_{0}\sim 14000~{\rm Mpc}$.
}
\label{fig1}
\end{figure*}

%%%%%%%%%%%%%%%%%%%%%%%%%%%%%%%%%%%%%%%%%%%%%%%%%%%%%%%%%%%%
\item \underline{\bf Step 4}: At any arbitrary momentum scale, $k$, the number of
e-foldings, $N(k)$, between the Hubble exit of the relevant modes, $k_\star$, and the end of inflation can be expressed as:
%
%\begin{widetext}
\be\begin{array}{llll}\label{eq8}
\displaystyle N(k) \approx  71.21 - \ln \left(\frac{k}{k_{0}}\right)  
+  \frac{1}{4}\ln{\left( \frac{V_{\star}}{M^4_{P}}\right) }
+  \frac{1}{4}\ln{\left( \frac{V_{\star}}{\rho_{end}}\right) }  
\\ \displaystyle ~~~~~~~~~~~~~~~~~~~~~~~~~~~~~~~~~~+ \frac{1-3w_{int}}{12(1+w_{int})} 
\ln{\left(\frac{\rho_{rh}}{\rho_{end}} \right)},
\end{array}\ee
where symbols are defined in Refs.~\cite{Choudhury:2013iaa,Choudhury:2014kma}. Now within the momentum interval, $k_{e}<k<k_{\star}$:
\be\begin{array}{llll}\label{eq9}
    \displaystyle \Delta N=N_{e}-N_{\star}=\ln\left(\frac{k_{\star}}{k_{e}}\right)\approx\ln\left(\frac{a_{\star}}{a_{e}}\right).
   \end{array}\ee
which can be recast as:
\be\begin{array}{llll}\label{eq10}
    \displaystyle \frac{k_{e}}{k_{\star}}\approx\frac{a_{e}}{a_{\star}}=e^{-\Delta N}
   \end{array}\ee
within this interval sub-Plankian field excursion $|\Delta\phi|<M_{p}$ implies that, $$\left|\frac{\Delta N V^{'}(\phi_0)M_{p}}{V(\phi_0)}\right|<1.$$
 For an example, if $|\Delta\phi|\sim {\cal O}(10^{-1}~M_{p})<M_{p}$ then within $\Delta N=17$ e-foldings we get roughly
 $\left|\frac{\Delta N V^{'}(\phi_0)M_{p}}{V(\phi_0)}\right|\sim {\cal O}(10^{-1}~M_{p})<M_{p}$. {\bf \red This further proves that the claim made in Ref.~\cite{Antusch:2014cpa} is incorrect.
Whatever approach one follows for the analytical computation, either in momentum space or in term of number of e-foldings, we always obtain the same order of magnitude 
as far as integration of Eq.~(\ref{eq7}) is concerned. }
Further using Eq~(\ref{eq10}) in Eq~(\ref{eq7}), we obtain: 
\be\begin{array}{llll}\label{eq11}
    \displaystyle \int^{{ k}_{\star}}_{{k}_{e}}\frac{dk}{k}\sqrt{\frac{r({k})}{8}}\\
\displaystyle=\sqrt{\frac{r(k_{\star})}{8}}\left[\left(2-\frac{a}{2}+\frac{b}{2}-\frac{c}{2}\right)\left[1-e^{-\Delta N}\right]
\right.\\ \left.\displaystyle ~~~~~~~~~~~~~~~~~ +\left(\frac{a}{2}-\frac{b}{2}+\frac{c}{2}-1\right)\Delta N~ e^{-\Delta N} 
\right.\\ \left.\displaystyle ~~~~~~~~~~~~~~~~~ -\left(\frac{b}{4}-\frac{c}{4}\right)(\Delta N)^{2}~ e^{-\Delta N} 
\right.\\ \left.\displaystyle ~~~~~~~~~~~~~~~~~~~~~~~~~~~~~~~~~~~~+\frac{c}{12}(\Delta N)^{3}~ e^{-\Delta N}\right].
   \end{array}\ee

\item \underline{\bf Step 5}: Now we substitute Eq~(\ref{eq11}) in Eq~(\ref{eq3}), and  we obtain our desired result:
\be\begin{array}{llll}\label{eq12}
    \displaystyle \sqrt{\frac{r(k_{\star})}{8}}\left|\left(\frac{a}{2}-\frac{b}{2}+\frac{c}{2}-2\right)\left[1-e^{-\Delta N}\right]
\right.\\ \left.\displaystyle ~~~~~~~~~~~~~~~~~ -\left(\frac{a}{2}-\frac{b}{2}+\frac{c}{2}-1\right)\Delta N~ e^{-\Delta N} 
\right.\\ \left.\displaystyle ~~~~~~~~~~~~~~~~~ +\left(\frac{b}{4}-\frac{c}{4}\right)(\Delta N)^{2}~ e^{-\Delta N} 
\right.\\ \left.\displaystyle ~~~~~~~~~~~~~~~~~~~~~~~~~~~~~~~~~~~~-\frac{c}{12}(\Delta N)^{3}~ e^{-\Delta N}\right|\approx \displaystyle\frac{|\Delta\phi|}{M_p}\,
   \end{array}\ee
In order the check the contributions from each term, let us explicitly write down the factors 
$a$, $b$ and $c$ in terms of the slow-roll parameters (see Refs.~\cite{Choudhury:2013iaa,Choudhury:2013jya} for the details where all the inflationary observables are explicitly 
written in terms of the slow-roll parameters $\epsilon_{V},\eta_{V},\xi^{2}_{V}$ and $\sigma^{3}_{V}$), as:
\be\begin{array}{lll}\label{eq13}
    \displaystyle a\approx \left[\frac{r(k_{\star})}{4}-2\eta_{V}(k_{\star})-4\left(2{\cal C}_{E}+\frac{1}{3}\right)\epsilon_{V}(k_{\star})\eta_{V}(k_{\star})
\right.\\ \left. \displaystyle ~~~~~~~~~~~~~~~~~
~~~~~~~~~~~~~~~~-4\left(6{\cal C}_{E}+\frac{11}{3}\right)\epsilon^{2}_{V}(k_{\star})
\right.\\ \left.\displaystyle~~~~~~~~~~~~~~~~~~~~~~+2{\cal C}_{E}\xi^{2}_{V}(k_{\star})-\frac{2}{3}\eta^{2}_{V}(k_{\star})+\cdots\right],\\
    \displaystyle b\approx \left[16\epsilon^{2}_{V}(k_{\star})-12\epsilon_{V}(k_{\star})\eta_{V}(k_{\star})+2\xi^{2}_{V}(k_{\star})+\cdots\right],\\
\displaystyle c\approx \left[-2\sigma^{3}_{V}+\cdots\right],
   \end{array}\ee
where $``\cdots''$ involves higher powers of the slow-roll contributions which are negligibly small in the leading order to hold the convergence 
criteria of the Taylor series mentioned 
in Eq~(\ref{eq1}). 

Substituting Eq~(\ref{eq13}) in Eq~(\ref{eq11}), we further obtain:

\be\begin{array}{llll}\label{eq14}
    \displaystyle 2\times\sqrt{\frac{r(k_{\star})}{8}}\left|\left\{\frac{r(k_{\star})}{16}-\frac{\eta_{V}(k_{\star})}{2}-1%\right]
-\left(6{\cal C}_{E}+\frac{23}{3}\right)\epsilon^{2}_{V}(k_{\star})
 \right.\right.\\ \left. \left.\displaystyle ~~~~~~~~~~~~~~~~-\frac{\eta^{2}_{V}(k_{\star})}{6} +\left({\cal C}_{E}-1\right)\frac{\xi^{2}_{V}(k_{\star})}{2}\right.\right.\\ \left.\left.~~~~~~~~~~
\displaystyle-\left(2{\cal C}_{E}-\frac{8}{3}\right)\eta_{V}(k_{\star})\epsilon_{V}(k_{\star})\right.\right.\\ \left.\left.~~~~~
\displaystyle-\frac{\sigma^{3}_{V}(k_{\star})}{2}+\cdots\,\right\}\left[1-e^{-\Delta N}\right]
\right.\\ \left.\displaystyle ~~~~~~~~ -\left\{\frac{r(k_{\star})}{16}-\frac{\eta_{V}(k_{\star})}{2}-\frac{1}{2}%\right]
-\left(6{\cal C}_{E}+\frac{23}{3}\right)\epsilon^{2}_{V}(k_{\star})
 \right.\right.\\ \left. \left.\displaystyle ~~~~~~~~~-\frac{\eta^{2}_{V}(k_{\star})}{6} +\left({\cal C}_{E}-1\right)\frac{\xi^{2}_{V}(k_{\star})}{2}\right.\right.\\ \left.\left.~~~~~~~~~~
\displaystyle-\left(2{\cal C}_{E}-\frac{8}{3}\right)\eta_{V}(k_{\star})\epsilon_{V}(k_{\star})\right.\right.\\ \left.\left.~~~~~~~
\displaystyle-\frac{\sigma^{3}_{V}(k_{\star})}{2}+\cdots\,\right\}\Delta N~ e^{-\Delta N}
\right.\\ \left. \displaystyle
+\left\{2\epsilon^{2}_{V}(k_{\star})-\frac{3}{2}\epsilon_{V}(k_{\star})\eta_{V}(k_{\star})
\right.\right.\\ \left.\left. \displaystyle+\frac{\xi^{2}_{V}}{4}+\frac{\sigma^{3}_{V}}{4}\left[1+\frac{2}{3}\Delta N\right]\right\}(\Delta N)^{2}~ e^{-\Delta N}\right|\\
\displaystyle~~~~~~~~~~~\approx \displaystyle\frac{|\Delta\phi|}{M_p}\,
   \end{array}\ee

\item \underline{\bf Step 6}:
Now within $\Delta N=17$ e-foldings from Eq~(\ref{eq10}), we obtain:
\be\begin{array}{llll}\label{eq10a}
    \displaystyle \frac{k_{e}}{k_{\star}}\approx\frac{a_{e}}{a_{\star}}=e^{-\Delta N}=e^{-17}=4.1\times 10^{-8}.
   \end{array}\ee
For an example, let us fix the momentum scale at, $k_{\star}=0.002~{\rm Mpc}^{-1}$, at the pivot scale, and then using Eq~(\ref{eq10a}), we get,
$k_{e}=8.2\times 10^{-11}~{\rm Mpc}^{-1}$. 

In this context the scalar power spectrum, spectral tilt, running of the tilt and running of the running of tilt for the scalar perturbations can be written as:
\be\begin{array}{llll}\label{power}
    \displaystyle P_{S}(k)=P_{S}(k_{\star})\left(\frac{k}{k_{\star}}\right)^{n_{S}(k_{\star})-1+\frac{\alpha_{S}(k_{\star})}{2}\ln\left(\frac{k}{k_{\star}}\right)
+\frac{\kappa_{S}(k_{\star})}{6}\ln^{2}\left(\frac{k}{k_{\star}}\right)}\,
   \end{array}\ee
\be\begin{array}{llll}\label{tilt}
    \displaystyle n_{S}(k)=n_{S}(k_{\star})+\alpha_{S}(k_{\star})\ln\left(\frac{k}{k_{\star}}\right)
+\frac{\kappa_{S}(k_{\star})}{2}\ln^{2}\left(\frac{k}{k_{\star}}\right)\,
   \end{array}\ee
\be\begin{array}{llll}\label{run1}
    \displaystyle \alpha_{S}(k)=\alpha_{S}(k_{\star})
+\kappa_{S}(k_{\star})\ln\left(\frac{k}{k_{\star}}\right)\,
   \end{array}\ee
\be\begin{array}{llll}\label{run2}
    \displaystyle \kappa_{S}(k)\approx\kappa_{S}(k_{\star})\,.
   \end{array}\ee
Similar relations can be obtained for tensor modes also.
At $k=k_{e}$, using Eq~(\ref{eq10}) in Eq~(\ref{power}), we obtain:
\be\begin{array}{llll}\label{power1}
    \displaystyle P_{S}(k_{e})=P_{S}(k_{\star})\left(\frac{k_{e}}{k_{\star}}\right)^{n_{S}(k_{\star})-1+\frac{\alpha_{S}(k_{\star})}{2}\ln\left(\frac{k_{e}}{k_{\star}}\right)
+\frac{\kappa_{S}(k_{\star})}{6}\ln^{2}\left(\frac{k_{e}}{k_{\star}}\right)}\,\\
\displaystyle =P_{S}(k_{\star})\left(e^{-\Delta N}\right)^{n_{S}(k_{\star})-1+\frac{\alpha_{S}(k_{\star})}{2}\ln\left(e^{-\Delta N}\right)
+\frac{\kappa_{S}(k_{\star})}{6}\ln^{2}\left(e^{-\Delta N}\right)}\,
\\
\displaystyle =P_{S}(k_{\star})\left(e^{-\Delta N}\right)^{n_{S}(k_{\star})-1-\frac{\alpha_{S}(k_{\star}) \Delta N}{2}
+\frac{\kappa_{S}(k_{\star})(\Delta N)^{2}}{6}}\,
   \end{array}\ee
\be\begin{array}{llll}\label{tilt1}
    \displaystyle n_{S}(k_{e})=n_{S}(k_{\star})+\alpha_{S}(k_{\star})\ln\left(\frac{k_{e}}{k_{\star}}\right)
+\frac{\kappa_{S}(k_{\star})}{2}\ln^{2}\left(\frac{k_{e}}{k_{\star}}\right)\,\\
~~~~~~~~~\displaystyle =n_{S}(k_{\star})-\alpha_{S}(k_{\star})\Delta N
+\frac{\kappa_{S}(k_{\star})(\Delta N)^{2}}{2}\,
   \end{array}\ee
\be\begin{array}{llll}\label{run1a}
    \displaystyle \alpha_{S}(k_{e})=\alpha_{S}(k_{\star})
+\kappa_{S}(k_{\star})\ln\left(\frac{k_{e}}{k_{\star}}\right)\,\\
~~~~~~~~~\displaystyle =\alpha_{S}(k_{\star})
-\kappa_{S}(k_{\star})\Delta N\,
   \end{array}\ee
\be\begin{array}{llll}\label{run2}
    \displaystyle \kappa_{S}(k_{e})\approx \kappa_{S}(k_{\star})\,.
   \end{array}\ee
 Since the reconstruction technique studied in  Ref.~\cite{Choudhury:2014kma} demands the amplitude of the scalar power spectrum $P_{S}$,
spectral tilt $n_{S}$, running of the tilt $\alpha_{S}$, and the running of the running of tilt $\kappa_{S}$ at the pivot scale $k_{\star}(=0.002~{\rm Mpc}^{-1})$ 
perfectly fits with the present data from Planck, we take the central values of these observables, as quoted in \cite{Choudhury:2014kma}. 
Within 17 e-foldings, using Eq~(\ref{power1}), we yield:
\be\begin{array}{llll}\label{value}
P_{S}(k_{e})\approx 6.27\times 10^{-9}\times P_{S}(k_{\star}),\\
\displaystyle n_{S}(k_{e})\approx 4.4\times n_{S}(k_{\star}),\\
\displaystyle \alpha_{S}(k_{e})\approx 16.45\times \alpha_{S}(k_{\star}),\\
\displaystyle \kappa_{S}(k_{e})\approx \kappa_{S}(k_{\star}).\end{array}\ee
where  $k_{\star}=0.002~{\rm Mpc}^{-1}$ and $k_{e}=8.2\times 10^{-11}~{\rm Mpc}^{-1}$ within $\Delta N=17$.
In fig~(\ref{fig1}) we have explicitly shown the behaviour of the power spectrum. 
Within this 17 e-foldings, we have $e^{-\Delta N}=4.1\times 10^{-8}<<1$, for which the factor $\left[1-e^{-\Delta N}\right]\approx 1$,
$\Delta N~e^{-\Delta N}=6.9\times 10^{-7}$ and $(\Delta N)^{2}~e^{-\Delta N}=1.1\times 10^{-5}$. 
 
Also within the slow-roll regime the 
slow-roll parameters $\epsilon_{V}<<1,\eta_{V}<<1,\xi^{2}_{V}<<1$ and $\sigma^{3}_{V}<<1$ for which the co-efficient of $\Delta N~e^{-\Delta N}$
 and $(\Delta N)^{2}~e^{-\Delta N}$ are also very small
at the leading order. 

Further if we multiply this small contribution with $\Delta N~e^{-\Delta N}=6.9\times 10^{-7}$ and $(\Delta N)^{2}~e^{-\Delta N}=1.1\times 10^{-5}$ 
within 17 e-foldings of inflation
the total contribution is negligibly small compared to the co-efficient of $\left[1-e^{-\Delta N}\right]\approx 1$ within 17 e-foldings.

{\bf \red Now, let us point out another mistake committed by the authors in Ref.~\cite{Antusch:2014cpa}, which is even more serious. The Ref.~\cite{Antusch:2014cpa} 
claimed that we have neglected and underestimated the leading order contribution in, $[1-e^{-\Delta N}]\approx 1-(1-\Delta N+...)\approx \Delta N$, which is ${\cal O}(17)$ for 
$\Delta N=17$. Numerically this argument is grossly incorrect, since the truncation of $e^{-\Delta N}$ series is not feasible for a large
 exponent.}
 
For the cross check,  let us expand the term: $[1-e^{-\Delta N}]$, which will yield:
$$\Delta N-\frac{(\Delta N)^2}{2}+\frac{(\Delta N)^3}{6}-\frac{(\Delta N)^4}{24}+\cdots$$ This implies that for $\Delta N=17$ the higher contributions are even larger. So 
the trunction of $[1-e^{-\Delta N}]$ is not at all possible. To get a proper result, we would need to consider the the full expression of $[1-e^{-\Delta N}]$,
which is ${\cal O}(1)$ for $\Delta N=17$ e-foldings. {\bf \red This again proves that the claim in Ref.~\cite{Antusch:2014cpa} is completely incorrect. The authors in Ref.~\cite{Antusch:2014cpa}
didn't get the correct numerical result, since they had ignored the higher order larger terms in the series of: $[1-e^{-\Delta N}]$, and quoted their results only from the first 
term of the series i.e. $\Delta N$. }
 
We hope our clarification completely nullifies the claim made in Ref.~\cite{Antusch:2014cpa} regarding the issue of 
getting wrong result by a factor of $\sim10-30$ for most values of $a > b > c$.

\item \underline{\bf Step 7}: Using these facts, we can recast Eq~(\ref{eq14}) as:
\be\begin{array}{llll}\label{eq14}
    \displaystyle 2\times \sqrt{\frac{r(k_{\star})}{8}}\left|\left\{\frac{r(k_{\star})}{16}-\frac{\eta_{V}(k_{\star})}{2}-1%\right]
-\left(6{\cal C}_{E}+\frac{23}{3}\right)\epsilon^{2}_{V}(k_{\star})
 \right.\right.\\ \left. \left.\displaystyle ~~~~~~~~~~~~~~~~-\frac{\eta^{2}_{V}(k_{\star})}{6} +\left({\cal C}_{E}-1\right)\frac{\xi^{2}_{V}(k_{\star})}{2}\right.\right.\\ \left.\left.~~~~~~~~~~
\displaystyle-\left(2{\cal C}_{E}-\frac{8}{3}\right)\eta_{V}(k_{\star})\epsilon_{V}(k_{\star})\right.\right.\\ \left.\left.~~~~~
\displaystyle-\frac{\sigma^{3}_{V}(k_{\star})}{2}+\cdots\,\right\}\right|\approx \displaystyle\frac{|\Delta\phi|}{M_p}\,
   \end{array}\ee
where the denominators of $r(k_{\star})$ can be normalized according to upper bound of BICEP2 and Planck~\footnote{One can also verify that within the range of field excursion, $\Delta \phi\sim (10^{-1}M_{p})<M_{p}$, it is possible to generate large tensor modes, with $r\geq 0.1$. For an example, in the case of a high scale MSSM inflation, with $\eta_{V}(k_{\star})\sim {\cal O}(10^{-2})$, and
 $\Delta\phi\sim {\cal O}(10^{-1}M_{p})<M_{p}$, it is possible to obtain:  $r\sim {\cal O}(0.12-0.27)$.} (See the analogous expressions in 
Refs.~\cite{Choudhury:2013iaa} and \cite{Choudhury:2013jya}, where the prefactors and the denominators of $r(k_{\star})$ were adjusted according 
to the upper bound of BICEP2 and Planck data.).

\end{itemize}

We hope the detailed discussions are sufficient enough to prove that the sub-Planckian field excursion models can also generate large tensor modes
  with $r\geq 0.1$, while
completely falsifying the claims presented in Ref.~\cite{Antusch:2014cpa}. We also hope that after this clarification the readers can appreciate that there is indeed a 
possibility of getting large tensor to scalar ratio, or large tensor modes from $\Delta\phi<M_{p}$ by violating the well known {\it Lyth bound}, and satisfy all the current observational 
constraints as explained in our Ref.~\cite{Choudhury:2013iaa}.

{\it Acknowledgements:} The authors have benefitted from the discussions with Seshadri Nadathur and Shaun Hotchkiss.

%%%%%%%%%%%%%%%%%%%%%%%%%%%%%%%%%%%%%%%%%%%%%%%%%%%%%%%%%%%%%%%%%%%%%%%%%%%%%%%%%%%%%%%%%%%%%%%%%%%%%%%%%%%%%%%%%%%%%%%%%%%%%%%%%%%%%%%%%%%%%%%%%%%%%%%%%%%%%%%%%%%%%%%%%%%%%%%%%%%%%%%%%%%
%%%%%%%%%%%%%%%%%%%%%%%%%%%%%%%%%%%%%%%%%%%%%%%%%%%%%%%%%%%%%%%%%%%%%%%%%%%%%%%%%%%%%%%%%%%%%%%%%%%%%%%%%%%%%%%%%%%%%%%%%%%%%%%%%%%%%%%%%%%%%%%%%%%%%%%%%%%%%%%%%%%%%%%%%%%

%\end{thebibliography}


\begin{references}

\bibitem{Choudhury:2013iaa}
  S.~Choudhury and A.~Mazumdar,
  %``An accurate bound on tensor-to-scalar ratio and the scale of inflation,''
  Nucl.\ Phys.\ B {\bf 882} (2014) 386
  [arXiv:1306.4496 [hep-ph]].
  %%CITATION = ARXIV:1306.4496;%%
  %15 citations counted in INSPIRE as of 11 Apr 2014

\bibitem{Choudhury:2014kma}
  S.~Choudhury and A.~Mazumdar,
  %``Reconstructing inflationary potential from BICEP2 and running of tensor modes,''
  arXiv:1403.5549 [hep-th].
  %%CITATION = ARXIV:1403.5549;%%
  %8 citations counted in INSPIRE as of 11 Apr 2014k

\bibitem{Antusch:2014cpa}
  S.~Antusch and D.~Nolde,
  %``BICEP2 implications for single-field slow-roll inflation revisited,''
  arXiv:1404.1821 [hep-ph].
  %%CITATION = ARXIV:1404.1821;%%

\bibitem{Choudhury:2013jya}
  S.~Choudhury, A.~Mazumdar and S.~Pal,
  %``Low & High scale MSSM inflation, gravitational waves and constraints from Planck,''
  JCAP {\bf 1307} (2013) 041
  [arXiv:1305.6398 [hep-ph]].
  %%CITATION = ARXIV:1305.6398;%%
  %15 citations counted in INSPIRE as of 11 Apr 2014

\end{references}
\end{document}